# Graphene-carbon nanotube composites

E.F.Sheka[1] and L.A.Chernozatonskii[2]

[1] Peoples' Friendship University of the Russian Federation, 117923 Moscow, Russia;
sheka@icp.ac.ru

[2] Institute of Biochemical Physics RAS, 19991 Moscow, Russia

sheka@icp.ac.ru

**Abstract**. The formation of graphen-nanotube composites addresses a few basic problems. First, both partners are good donors and acceptors of electrons, which significantly complicates the intermolecular interaction between them leading to a two-well shape of the ground state energy term. The second problem concerns odd-electron character of the components. Similarly to high aromatics and fullerenes, much larger C-C distances provide a considerable weakening of odd electrons interaction in nanotubes and graphene that necessitates taking the configurational interaction of odd electrons into account. Avoiding a severe complication, the broken spin-symmetry approach makes the problem feasible. Moreover, unrestricted broken-symmetry Hartree-Fock approach possesses a unique sensitivity in revealing enhanced chemical activity of the species caused by their partial radicalization in terms of atomic chemical susceptibility. The chemical susceptibility profiles along the tube and across their body as well as over graphene sheets form the ground of computational synthesis of graphen-nanotube composites in due course of the relevant addition reactions and make it possible to select two main groups of the composites, conditionally called hammer and cutting-blade structures. The final product will depend on whether both components of the composition are freely accessible or one of them is fixed. Thus, in diluted solutions where the first requirement is met, one can expect the formation of the multi-addend cutting-blade composites. Oppositely, when either nanotubes or graphene sheets are fixed on some substrates, the hammer composites will be formed. A particular "cradle" composite is suggested for an individual graphene sheet to be fixed by a pair of nanotubes.

**Keywords:** carbon nanotubes, graphene, unrestricted broken spin-symmetry Hartree-Fock approach, chemical susceptibility, computational synthesis, hammer composites, cutting-blade composites

## 1. Introduction. Composites in view of intermolecular interaction

Recently started the manufacturing of nanocarbon-based composite materials pursues well-defined goals to provide the best conditions for the exhibition and practical utilization of extraordinary thermal, mechanic, electronic, and chemical properties of the nanocarbons. Obvious success in achieving the goal in the case of carbon nanotubes (CNTs) [1-5] and graphene [6] dissolved in different polymers points to great perspectives of a new class of composites and their use in a variety of applications. These nanocarbons (dopants) homogeneously dispersed in polymer matrices at very low concentration exhibit electrical and mechanical properties at the level of meeting the requirements of their exploitation at practice. Simultaneously, not only the preservation of the properties of individual tubes or graphene sheets, but a considerable enhancement of the latter has been observed.

Since low-concentration solutions of individual CNTs and graphene sheets can be obtained [1-6], one can put a question what can we expect when both CNTs and graphene are dissolved simultaneously? From the basic standpoint the question is addressed to the problem of the intermolecular interaction (IMI) between the dopant individuals, both related to the same species and to a mixture of them, as well as between dopants and a matrix. In both cases, the IMI peculiarities are connected with a considerable contribution of the donor-acceptor (DA) interaction since both CNTs and graphene are good donors and acceptors of electron while some matrices are good electron donors. In what follows we will concentrate on the first problem leaving the interaction between dopants and matrices outside. The latter is undoubtedly rather important since it might cause the above mentioned enhancement of the dopant properties and deserves a particular consideration. Intuitively, it should be similar to that disclosed for fullerene solutions [7, 8].

As for the interaction between dopants, due to similarity of both ionization potentials $I$ and electron affinities $\varepsilon$ of the species the potential energy surfaces (IMI terms) $E_{int}(r,R)$ ($r$ and $R$ denote intra- and intermolecular coordinates, respectively) for the ground state of pairs CNT+CNT (*I+I*); grph+grph (*II+II*), and CNT+grph (*I+II*) (*I* and *II* label pair partners) are well similar and posses a characteristic two-well structure as a function of intermolecular coordinates $R$ [9]. Generally, the shape of the IMI term of a pair consisting of electron donor *A* and electron acceptor *B* depends on the energy gap $I_A - \varepsilon_B$ [9, 10]. In the case of the pairs mentioned above, the relevant gap $I_A - \varepsilon_B$ provides the IMI term shown in the Fig.1. The term is a composition of terms $E_{int}(A^+B^-)$ and $E_{int}(A^0B^0)$, where the former describes the interaction between ions, which leads to their coupling at point $R^{(+\ -)}$ while the latter relates to neutral moieties coupled at point $R^{(0\ 0)}$. The formation of a stationary product *AB* at point $R^{(+\ -)}$ is accompanied by the creation of "intermolecular" chemical bonds between *A* and *B* partners. Oppositely, widely spaced neutral moieties form a charge transfer (CT) complex *A+B* in the vicinity of point $R^{(0\ 0)}$ [9, 10]. In ensembles of the above partners, the availability of a pair wise interaction minimum at point $R^{(0\ 0)}$ leads to the clusterization of the dopants in solutions of the $[A]_n$, $[B]_m$, or $A_k[B]_l$ types which actually takes place in the case of fullerenes [7, 8, 11] where solvent plays the role of partner *B*. Electronically, the formed nanosize clusters provide good conditions for electron properties space confining that might be responsible for the enhancement of the dopant properties mentioned in [1-6]. Leaving a detailed consideration of the problem for the future, we will concentrate on the formation of covalently bound products *AB* in the vicinity of point $R^{(+\ -)}$.

Three chemical additive reactions can be expected in mixed solutions of the partners that lead to the formation of chemically bound products, namely:

CNT composites $(I)_n$;

Nanographene (Ngr) composites $(II)_m$;

mixed CNT-Ngr composites $(I)_k(II)_l$.

For simplicity, following the tendency exhibited in [5], we extend the usage of the term "composites" from diluted solutions of nanocarbons in polymers to the nanocarbon products of the above chemical reactions $(I)_n$, $(II)_m$, and $(I)_k(II)_l$

First two reactions concern coalescence of CNTs and graphene sheets which is experimentally observed quite often and whose description can be generally understood on the basis of the detailed consideration of the fullerene $C_{60}$ dimerization [10]. The formation of mixed composites has become known just recently [12]. The obtained composite cannot be attributed to the previously mentioned cases [1-6] characterized by low concentration of the dopants. However, the finding has disclosed that the formation of mixed CNT-graphene constructions is possible. This makes the expectation of CNT-Ngr compositions in diluted solutions quite promising. Nothing is known about the shape and properties of such composites. The current paper presents a quantum chemical (QCh) view on the matter. Basing on atomically matched chemical susceptibility (CS) of both CNTs and graphene, we suggest a set of $(I)_k(II)_l$ ($k$= 1,2; $l$ = 1, 2) composites exhibiting the most probable structures to be experimentally synthesized.

## 2. Grounds for the computational synthesis of $(I)_k(II)_l$ composites

To simulate the formation of $(I)_k(II)_l$ product, one has to build a starting configuration consisting of components *I* and *II*. As seen in Fig.1, the intermolecular C-C distances at the reactive spot should be shorter than $R^{(0\ 0)}$ but longer than $R^{(+\ -)}$. Thoroughly analyzed in [10] by calculating the dimerization barrier for fullerene $C_{60}$, the distance should be of 1.9-1.6Å. Since in the composites under study intermolecular chemical bonds are formed by carbon atoms, conclusions obtained for fullerenes should be valid for the current case as well. In what follows the C-C starting distances are taking at 1.8-1.9Å.

A choice of the atomically matched reactive spot is the most complicated part of the simulation. To solve the problem we use atomic chemical susceptibility (ACS) maps discussed in details for single-walled CNTs (SWCNTs) in [13] and for Ngrs in [14]. A typical pattern of the ACS profile along an arbitrary SWCNT is shown in Fig.2 exemplified by data for (10,10) tube [13]. A thorough analysis of the peculiarities of the SWCNT ACS profile allowed for making a few conclusions concerning addition reactions to be expected:

- The space of chemical reactivity of SWCNTs coincides with the coordinate space of their structures whilst different for particular structure elements. This both complicates and facilitates chemical reactions involving the tubes depending on a particular reaction goal.
- Local additions of short-length addends (involving individual atoms, simple radical and so forth) to any SWCNT are the most favorable at open empty ends, both armchair and zigzag ones, the latter more effective. Following these places in activity are cap ends, defects in the tube sidewall, and sidewall itself. The reactivity of the latter is comparable with the highest reactivity of fullerene atoms.
- Chemical contacts of SWCNTs with spatially extended reagents (graphene sheets) can occur in three ways when the tube is oriented either normally or parallel to the surface and when graphene acts as a "cutting blade".
- Addition reactions with the participation of multi-walled CHTs will proceed depending on the target atoms involved. If empty open ends of the tubes are main targets, the reaction will occur as if one deals with an ensemble of individual

SWCNTs. If sidewall becomes the main target the reaction, output will depend on the accessibility of inner tubes additionally to the outer one.

A concentrated view on the reactivity of atoms of a rectangular Ngr [14] is presented in Fig.3. Similarly to the above, a thorough study of graphene has shown that

- Any chemical addend will be attached to Ngr zigzag edges first of all, both hydrogen terminated and empty.
- Slightly different by activity armchair edges of non-terminated Ngrs compete with zigzag edges.
- Chemical reactivity of inner atoms does not depend on the edge termination and is comparable with that of CNT sidewall and fullerenes thus providing a range of addition reactions at Ngr surface.

Calculations in the current study were performed on the platform of unrestricted broken spin-symmetry Hartree-Fock approach [15-18] whose weak and strong points as well as undoubtedless preference over unrestricted broken-symmetry DFT approach when applying to quantitative description of the odd-electron structure of CNTs and graphene are discussed in [13, 14]. A semi-empirical implementation of the approach in terms of AM1 version of the CLUSTER-Z1 program was exploited.

## 3. Computational synthesis of $(I)_k(II)_l$ composites

Empirical chemical procedures that can be suggested for mixed CNTs and Ngrs solutions to be formed [1-6], definitely provide removing hydrogen terminators from both components. That is why SWCNTs with empty ends and Ngrs with non-saturated edges will be mainly considered. Oppositely to convenient reagents, for which spots of chemical reactivity are locally concentrated, the latter for both CNTs and Ngrs is space-distributed. Consequently, there might be endless number of compositions formed by the components. However, basing on the analysis of SWCNTs and Ngrs ACS profiles presented in Fig.2 and Fig.3, it is possible to suggest two basic groups of composites below referred to as *hammer* and *cutting blade structures*. The former follows from the fact that empty ends of SWCNTs are the most chemically active so that the tubes might be willingly attached to any Ngr sheet forming a "hammer handle". In its turn, the sheet surface is reactive enough to provide chemical bonding with the tube. As for the second type of structures, this is a result of exclusive chemical reactivity of both zigzag and armchair edges of non-terminated Ngr, so that the latter can touch a SWCNT sidewall that is chemically active as well, as a blade. In what follows we shall restrict ourselves by composites with $k=1, 2$ and $i=1, 2$ and consider a few examples of the $(I)_{1,2}(II)_{1,2}$ compositions of both kinds. Besides the equilibrium structures that are obtained in due course of the structure optimization when looking for the energy minimum, the obtained composites will be characterized by the coupling energy $E_{cpl}$ per one intermolecular C-C bond formed at the interface determined as

$$E_{cpl} = (\Delta H_{cps} - k\Delta H_{SWCNT} - i\Delta H_{Ngr})/n_{C-C} \qquad (1)$$

where $\Delta H_{cps}, \Delta H_{SWCNT}$, and $\Delta H_{Ngr}$ present the heats of formation of the composite, nanotube, and Ngr, respectively, and $n_{C-C}$ is the number of intermolecular C-C bonds formed. The obtained data are listed in Table 1.

## 3.1. Hammer $(I)_{1,2}(II)_{1,2}$ composites

We have chosen a few composites to exhibit general tendencies that govern the formation of the SWCNT-Ngr interface. A summarized view on equilibrium structures of the studied composites is presented in Fig.4. Each structure is nominated by a combined notation like I*a*, I*b* and so forth to simplify description of the data as well as their presentation in Table 1.

**I**. A fragment of (4,4) SWCNT with both open empty ends is normally oriented to Ngr (7,7) sheet at starting distance of 1.8Å (I*a*). Seven intermolecular C-C bonds of 1.51Å in length are formed joining the tube with the sheet after the structure optimization (I*b*). As seen from Table 1, the corresponding coupling energy is quite large pointing to a strong connection at the interface. Additionally to the energy of the bond formation, the coupling energy involves the deformation energy of both SWCNT and Ngr caused by the reconstruction of the electron configuration of the interface carbon atoms from $sp^2$ to $sp^3$. However, when this reconstruction touches mainly upon the structure of the tube end, the Ngr is significantly deformed as a whole that leads to the transformation of a flat "roof" of the starting composition in I*a* to a "Chinese pagoda" pattern when the structure is equilibrated (I*b*). However, when the second Ngr sheet is added to the first one of the starting composition in Fig 4.I*a* at 3.35 Å above, the intermolecular interaction between two sheets evidently smoothes the deformation of the first one, while the second is only slightly deformed (I*c*). At the same time, the coupling energy of the tube to the sheet pair remains big and decreases rather slightly. Evidently, coupling three-layer graphene stack with the nanotubes will be strong as well while the deformation of the upper sheets will be practically negligible. As might be expected, the obtained results are in line with VASP calculations of the connection of the graphene sheet stack with the SiC substrate [19] promoting the growth of epitaxial graphene [20].

**II**. When the tube is placed parallel to the sheet plane (II*a*) at the same distance as in case **I**, the equilibrium structure occurs to depend on whether the tube open ends are either empty or terminated (by hydrogens in our case). In the first case, the tube and the sheet attract each other willingly and seven newly formed intermolecular C-C bonds provide the tight connection between the partners (II*b*). When tracing subsequent steps of the joining (optimization), one can see that the coupling starts at the tube ends by the formation of a single bond at first and then a pair of the C-C bonds at each end. Afterwards, these bonds play the role of the strops of gymnastic rings that pull the tube body to the sheet. Oppositely, when the tube ends are hydrogen terminated, no intermolecular C-C bonds are formed (II*c*) and the total energy coupling energy becomes repulsive (See Table 1). However, even under this repulsive interaction the sheet is deformed showing a tendency to wrapping up the tube.

**III**. Attachment of two SWCNTs to a single Ngr (III*a*) shows that the final result depends on the topological coherence between the tube projection and the benzenoid structure of the sheet (III*c*). The left joining in III*b* occurs evidently under better coherence than the right one. This causes the different number of the formed intermolecular C-C bonds, namely, eight in the first case and two in the second. Both attachments are accompanied with the sheet deformation that causes a remarkable roughening of the latter. The coupling energy is comparable with that for a single SWCNT attachment.

**IV**. The fragment of a double-wall CNT (DWCNT) in IV*a* consists of fragments of (4,4) and (9,9) SWCNTs with the same number of benzoid units along the vertical axis and open empty ends on both sides. However, since the periodicity of the Kekule-incomplete Clar-complete Clar networks [21] is slightly different in the two tubes, the fragment lengths do not coincide exactly. In due course of the optimization, the attachment of the joint fragment to the graphene sheet starts from the formation of intermolecular C-C bonds with either inner or outer tube depending on which is closer to the sheet. When opposite free ends of the tube are not fixed, the remaining fragment slides outwards, transforming the composition into a peculiar telescope system. When the free ends are fixed, both inner and outer fragments are joined to the sheet (IV*b*). The coupling energy given in Table 1 cannot be directly compared with that of SWCNT since it is affected by both the sheet deformation and the free end fixation. Nevertheless, it is large enough to provide a strong coupling between the graphene sheet and the DWCNT that explains a high stability of recently synthesized MWCNTs-graphene composite under conditions when one end of each MWCNT was fixed [12].

The performed calculations make it possible to conclude the following.

1. The normal attachment of an empty-end SWCNT to graphene sheet is energetically favourable.
2. The horizontal attachement of the tube is also possible whilst much weaker.
3. H-termination of the tube ends renders the horizontal attachment impossible and severely weakens the normal one.
4. Both multiple normal attaching of SWCNTs as well as a single and multiple attaching of a DWCNT are energetically favourable and graphene sheets can be easily fixed over tubes in case their open ends are empty. This conclusion is in a perfect consent with experimental observation presented in [12].
5. Graphene sheets are extremely structure flexible and even a weak intermolecular interaction causes a loss of the sheet flatness.

## 3.2. Cutting-blade $(I)_{1,2}(II)_{1,2}$ composites

Two SWCNT fragments, namely (8,0) and (4,4), as well as Ngr (7,7) were chosen to demonstrate typical compositions to be formed in this case. At start each time, the Ngr edge was oriented parallel to the cylinder axe in the vicinity of SWCNT along a line of sidewall atoms in such a way to maximize the number of expected intermolecular C-C bonds. Since ACS distribution over cross section atoms of both tubes is well homogeneous [13], there is no line selectivity in this case. As for Ngrs, zigzag and armchair edges of NGR with empty edges are comparable (see Fig.3) whilst somewhat different. Due to this, two NGR orientations with respect to the tube sidewall were examined.

Equilibrated structures of composites based on (8,0) SWCNT are shown in Fig. 5.

**V**. The formation of either zigzag (V*a*) or armchair (V*b*) attached monocomposites (*monoderivatives*) is followed by the creation of 8 intermolecular C-C bonds at the interface in both cases. However, the two composites differ by the coupling energy (Table 1), therewith V*b* is more energetically profitable in spite of the zigzag edges ACS is slightly higher. The difference is a consequence of the interface different structure that causes different deformation energies (positive by sign), which is obviously bigger for composite V*a*.

**VI**. To look for a proper spot for the attachment of the second Ngr, let us analyze the ACS profile over the cross section of the tube body of both composites. The relevant distributions VI*a* and VI*b* occurred quite similar keeping the same view in all cross sections along the tube in both cases. It should be noted a vivid transformation of a circular cross section of a free tube into a drop-like one with the apex on atom A. It is obviously caused by the $sp^2 \rightarrow sp^3$ transformation at the point. In both cases, atom A, matched by small black circles in V*a* and V*b*, is involved into the line through which the first Ngr is attached to the tube. As shown [13], ACS is distributed over (8,0) SWCNT cross sections atoms quite homogeneously at almost constant level of $N_{DA}$ ~0.29. When one Ngr is attached, atom A is involved in the formation of an intermolecular C-C bond and, consequently, its $N_{DA}$ falls to zero. The remainder value of 0.01 can points to the reliability of the $N_{DA}$ values determination. Consequently, a redistribution of ACS over the cross section atom occurs that is clearly seen at VI*a* and VI*b* diagrams. According to basic concept of computational synthesis of odd electron systems with unpaired electrons [17, 18], the next attachment will be the most profitable at atoms with the highest $N_{DA}$ values. Looking at diagram VI*a*, we can conclude that such atoms are located on the right and on the left from atom A. However, attaching Ngr along the corresponding lines of atoms will meet sterical difficulties. The next suitable atom, matched as B, is located in three atoms from atom A. Symmetrically is located atom B` with similar characteristic. Between atoms B and B` there are more 7 atoms with the same ACS. As a result, the nearest attachment of the second NGR should occur along atom lines involving either B or B' atom, or any of seven atoms between them. For the first investigation we chose atom B. Similarly analyzed diagram VI*b* allows for concluding that the preferable place of the second Ngr attachment to the tube is the atom line involving atom C while line with atom B is the least preferable.

**VII** and **VIII**. Realizing the first of the above conclusions, dicomposite (*diderivative*) VII*a* was obtained. The creation of new 7 intermolecular C-C bonds accompanies the composite formation with the coupling energy averaged over 15 bonds presented in Table 1. Comparing composites V*a* and VII*a*, one can see that the coupling energy remains practically the same with a slight increasing in the latter case.

Composite VII*b* is a result of attaching the second Ngr along the line with atom B on diagram VI*b* while composite VIII is formed when the second Ngr is attached along the line with atom C. Eight new C-C bonds are formed in both cases with the coupling energy averaged over all formed C-C bonds for both composites shown in Table 1. According to the data, the two composites are characterized by large coupling energy, among which the glider-like composite VIII is the most energetically favorable. The energy difference for VII*b* and VIII composites could have been considered as an evidence of the preference of the addition reaction in place C against place B. However, one has to take into account the deformation energy which seems to be larger in case B.

The series of $(I)_1(II)_i$ composites is not restricted to *i*=1 and 2. To proceed further we have to determine the place of the next attachment, for which we have to look at the ACS profile over the cross section of the tube body after the second reaction. Naturally, the related diagrams differ from VI*a* and VI*b*. As follows from the calculations performed, in the case of composite VII*a*, atom C (see VI*a*) has the highest $N_{DA}$ value pointing the place of the third attachment. In the case of composites VII*b* and VIII, the corresponding places are marked by E and D(D`) (see VI*b*) respectively. Therefore, when the Ngr attaches the (8,0) SWCNT via zigzag edges, a series of multi additions will look as A(1) → B(2) → C(3) → and so on (see mapping in VI*a*). In the case of attaching via armchair edge, composite VIII is energetically preferable, so that the multi addition will follow a scheme A(1) → C(2) → D(3) or D`(3) → and so on (VI*b*) until "6-7-tooth Ngr gear" can be formed on the basis of the tube.

We have considered the simplest cases. In reality, addition reactions might be more complex since a mixed attachment cannot be excluded, when one Ngr attaches a tube via armchair edge while the other via zigzag one. The number of intermolecular C-C bonds formed at the interface will play an important role as well since this determines the total coupling energy. However, the presented scheme shows a definite way how any individual addition reaction can be traced.

Equilibrated structures of composites based on (4,4) SWCNT are shown in Fig. 6.

**IX**. Oppositely to composites based on (8,0) SWCNT, results of mono addition IX*a* and IX*b* show a clear preference of the zigzag attachment. Although the coupling energy seems to favor the second one, a great number of C-C bond formed as well as much weaker deformation at the interface evidently support the above preference.

**X**. The Ngr is attached to the tube along atom line involving atom A matched by small black circle in IX*a*. Similarly to the previous case, a drastic falling of the reactivity of atom A causes a severe distortion of the circular cross section of the tube that demonstrates ACS profile X passing through atom A. The cross section of the ACS distribution exhibits atoms B and C with the highest $N_{DA}$. Earlier we mentioned that placing the second Ngr in the exact neighborhood of the fixed atom may cause distortion of the structure due to sterical constrain. To check the expectation A,B and A,C double attachment were studied.

**XI**. As expected, A,B attachment causes a severe distortion at the interface fully destroying the tube structure (composite XI*a*). Important, that the composite is characterized by a large coupling energy as well as by a big number of intermolecular C-C bonds formed. Oppositely to the case, the formation of composite XI*b* does not cause any tube body destroying and dicomposite XI*b* is quite similar to VIII considered above for (8,0) SWCNT. The coupling energy is large; the number of the formed C-C bonds is big that all favors such composites formation in the diluted solutions. Nevertheless, composite XI*a* is obviously more energetically favorable, so that the real situation in a laboratory flask might be quite complex.

Analyzing ACS profile over cross section of composite XI*b* tube passing through atom marked by small black circle, one finds atoms B and B` (see X) with practically equal $N_{DA}$ values (B - 0.52 and B` - 0.49) that are twice bigger than the related values for other atoms. Under these conditions, the third attachment will occur along lines passing through these atoms. But those are in a direct neighborhood of atoms A and C so that one may expect a severe destroying the tube body structure similar to that of composite XI*a*. It cannot be excluded that this destroying is characteristic for small-diameter tubes only and that for large-diameter tubes a sequential addition of a number of Ngr will result in the formation of a multi-tooth gear as was in the case on (8,0) SWCNT. Both questions require further detailed investigations.

**XII**. A particular attention should be drawn to this cradle-like composite. As shown in Table 1, it is energetically stable. On the other hand, the calculations show one of possible ways of an individual graphene sheet fixation under conditions of the least perturbation of the sheet. Obviously, not (4,4) SWCNT but much larger tubes should be taken as supporters. Since ACS of SWCNTs does not much depend on the tube diameter [13], the cradle composite formation can be provided by any tubes, even different in diameter within the pair.

In spite of the doubtless exemplary of studied composites, the performed investigations allow for making the following general conclusions.

1. The formation of the hammer and cutting blade $(I)_k (II)_i$ composites is energetically favourable not only as mono addition of Ngr to the tube body and vice versa but as a multi-addend attachment as well.

2. A strong contact between the tube and Ngr is provided by the formation of intermolecular C-C bonds, number of which is comparable with the number of either tube end or Ngr edge atoms.

3. The contact strength is determined by both energy of newly formed C-C bond and their number. Optimization of the latter dictates a clear preference towards zigzag or armchair edges of the attaching Ngr depending on the tube configuration. Thus, (8,0) SWCNT (as all other members of the (m,0) family) prefers armchair contacts that maximizes the number of point contacts. In its turn, (4,4) SWCNT ( as well as other members of the (n,n) family) favours zigzag contacts due to the same reasons.

4. The total coupling energy between the Ngr addend and tube involves both the energy of C-C bond formed and the energy of deformation caused by the reconstruction of $sp^2$ configuration for the carbon atom valence electrons into $sp^3$ one. It can be thought that the latter depends on the tube diameter. However, the data are so far rather scarce and an extended investigation of the problem is needed.

5. In general, the coupling energy of cutting blade composites is much more than that of hammer ones that is important for a practical realization of the $(I)_k (II)_i$ composites production

6. The final product will depend on whether both components of the composition are freely accessible or one of them is rigidly fixed. Thus, in diluted solutions where the first requirement is met, one can expect the formation of cutting blade composites due to significant preference in the coupling energy. Oppositely, in gas reactors where often either CNTs or graphene sheets are fixed on some substrates, the hammer composites will be formed as it has been shown just recently [12].

**4. Conclusion**

The problem of SWCNT+graphen composites concerns a few basic problems due to extreme specificity of both components. Thus, they both are good donors and acceptors of electrons and this significantly complicates the intermolecular interaction leading to a two-well shape of the ground state energy term. This provides the formation of two composites, one of which consists of weakly interacting components located at comparatively large distance while the second is formed in the range of short interatomic distances and corresponds to strongly coupled composition. Both composites, expected to be drastically different by properties, may exist and should be differentiated. The first attempt to consider properties of strongly coupled composites is undertaken in the paper.

Next problem concerns odd-electron character of both components. Similarly to high aromatics and fullerenes, odd electrons of CNTs and graphene interact much weaker, than, say, in ethylene and benzene due to much larger C-C distances in the species. Consequently, a lot of nearly degenerate states appear in their energy spectrum due to which a theoretic description of their properties, particularly in singlet state, has to take configurational interaction into account. Avoiding severe computational difficulties, the broken spin-symmetry approach makes the problem feasible. Modern implementations of the approach in the form of either unrestricted Hartree-Fock scheme or unrestricted DFT were discussed in [13, 14] with particular attention to applicability of spin-contaminated solutions of both techniques to describe electronic properties of CNTs and graphene. While UBS DFT provides the determination of the singlet state energy that is closer to that of pure spin state, UBS HF demonstrates a unique sensitivity in revealing enhanced chemical activity of the species caused by their partial radicalization and provides a numerical presentation of the atomically matched chemical susceptibility of the species. Atomic chemical susceptibility profiles along the tube and across their body as well as over graphene sheets formed the ground of computational synthesis of CNTs+graphene composites in due course of the relevant addition reactions.

Two SWNT fragments presenting (n,n) and (m,0) families, namely, (4,4,) and (8,0) and a set of graphene sheets of different size were chosen to reveal general tendencies of the composite formation. Due to the fact that the space of chemical reactivity of both CNTs and graphene coincides with the coordinate space of their structures, addition reactions that lead to the composite formation are not local but largely extended in the space. This greatly complicates the construction of starting diads, triads, and more complex configurations of components making their number practically endless. However a thorough analysis of the ACS profiles of both components made it possible selecting two main groups of the composites, conditionally called hammer and cutting blade structures. The former follows from the fact that empty ends of SWCNTs are the most chemically active so that the tubes might be willingly attached to any Ngr forming a hammer handle. The latter is a consequence of exclusive chemical reactivity of both zigzag and armchair edges of non-terminated Ngr, so that Ngr can touch a SWCNT sidewall as a blade. As occurred, the coupling energy of cutting blade composites exceeds that of hammer ones that is important for a practical realization of the $(I)_k (II)_i$ composites production. The final product will depend on whether both components of the composition are freely accessible or one of them is fixed. Thus, in diluted solutions where the first requirement is met, one can expect the formation of the multi-addend cutting blade composites due to significant preference in the coupling energy. Among the latter, a particular "cradle" composite is suggested for an individual graphene sheet to be fixed by a pair of nanotubes. Oppositely, in gas reactors where often either CNTs or graphene sheets are fixed on some substrates, the hammer composites will be formed as it has been shown just recently [12].

**ACKNOWLEDGMENTS**. The work was supported by the Russian Foundation for Basic Research (grant 08-02-01096).

**Table 1**. Coupling energy at the interface of CNT+Ngr composites per one intermolecular C-C bond formed, *kcal/mol*

| Nomination [1] | Composites | $E_{cpl}$ |
|---|---|---|
| | Hammer $(I)_1(II)_{1,2}$ structures | |
| Ib | (4.4) SWCNT + (7,7) Ngr *(7)*[2] | -32.24 |
| Ic | (4.4) SWCNT + 2*(7,7) Ngr *(7)* | -28.13 |
| IIb | (4,4) SWCNT + (7,7) Ngr *(7)* | -11.30 |
| IIc | (4,4) SWCNT + (7,7) Ngr *(0)* | + 4.11[3] |
| IIIb | 2* (4,4) SWCNTs + (11,12) Ngr *(8+2)* | -42.64 |
| IV | (4,4 + 9,9) DWCNT+(11,12) Ngr *(8)* | -10.00 |
| | Cutting blade $(I)_{1,2}(II)_{1,2}$ structures | |
| Va | (8,0) SWCNT + (7, 7) Ngr z-*(8)*[4] | -52.49 |
| Vb | (8,0) SWCNT + (7, 7) Ngr a-*(8)* | -66,28 |
| VIIa | (8,0) SWCNT + 2*(7, 7) Ngr zz-*(15)* | -54.58 |
| VIIb | (8,0) SWCNT + 2*(7, 7) Ngr aa-*(16)* | -54.07 |
| VIII | (8,0) SWCNT + 2*(7, 7) Ngr aa-*(16)* | -62.43 |
| IXa | (4,4) SWCNT + (7, 7) Ngr z-*(7)* | -70.75 |
| IXb | (4,4) SWCNT + (7, 7) Ngr a-*(4)* | -77.84 |
| XIa | (4,4) SWCNT + 2*(7, 7) Ngr zz-*(14)* | -118.97 |
| XIb | (4,4) SWCNT + 2*(7, 7) Ngr zz-*(16)* | -83.50 |
| XII | 2*(4,4) SWCNT + (7, 7) Ngr zz-*(13)* | -69.86 |

[1] See corresponding structures in Figs. 4-6.

[2] The figure in brackets indicates the number of intermolecular C-C bonds formed

[3] The total interaction energy (see text)

[4] z(a)-(*n*) and zz(aa)-(*n*) indicate zigzag (armchair) single or double Ngr attachment to the tubes sidewall, respectively, accompanied by the formation of *n* intermolecular C-C bonds.

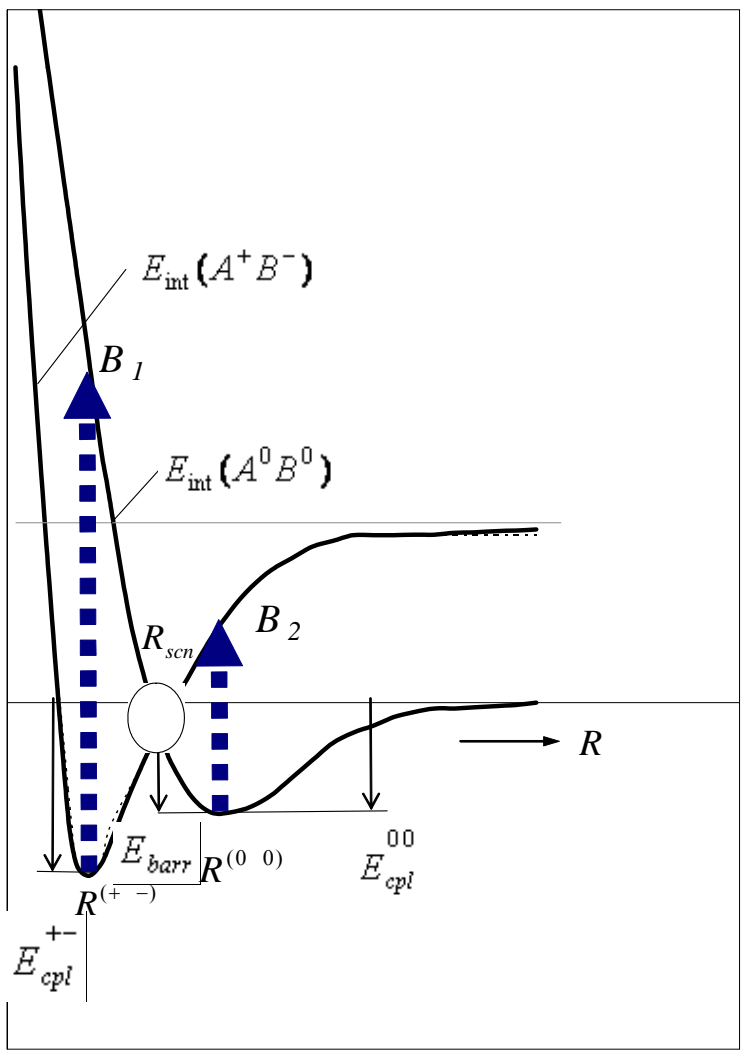

**Figure 1**. Cross section of the potential energy surface (IMI term) of a DA binary system along an arbitrary intermolecular coordinate. $E_{cpl}^{+-}$ and $E_{cpl}^{00}$ are coupling energies of the formation of chemically bound product *AB* and charge transfer complex *A+B*, respectively. $E_{barr}$ is barrier energy for the transition from *A+B* to *AB*. Arrows $B_1$ and $B_2$ depict optical electronic transitions.

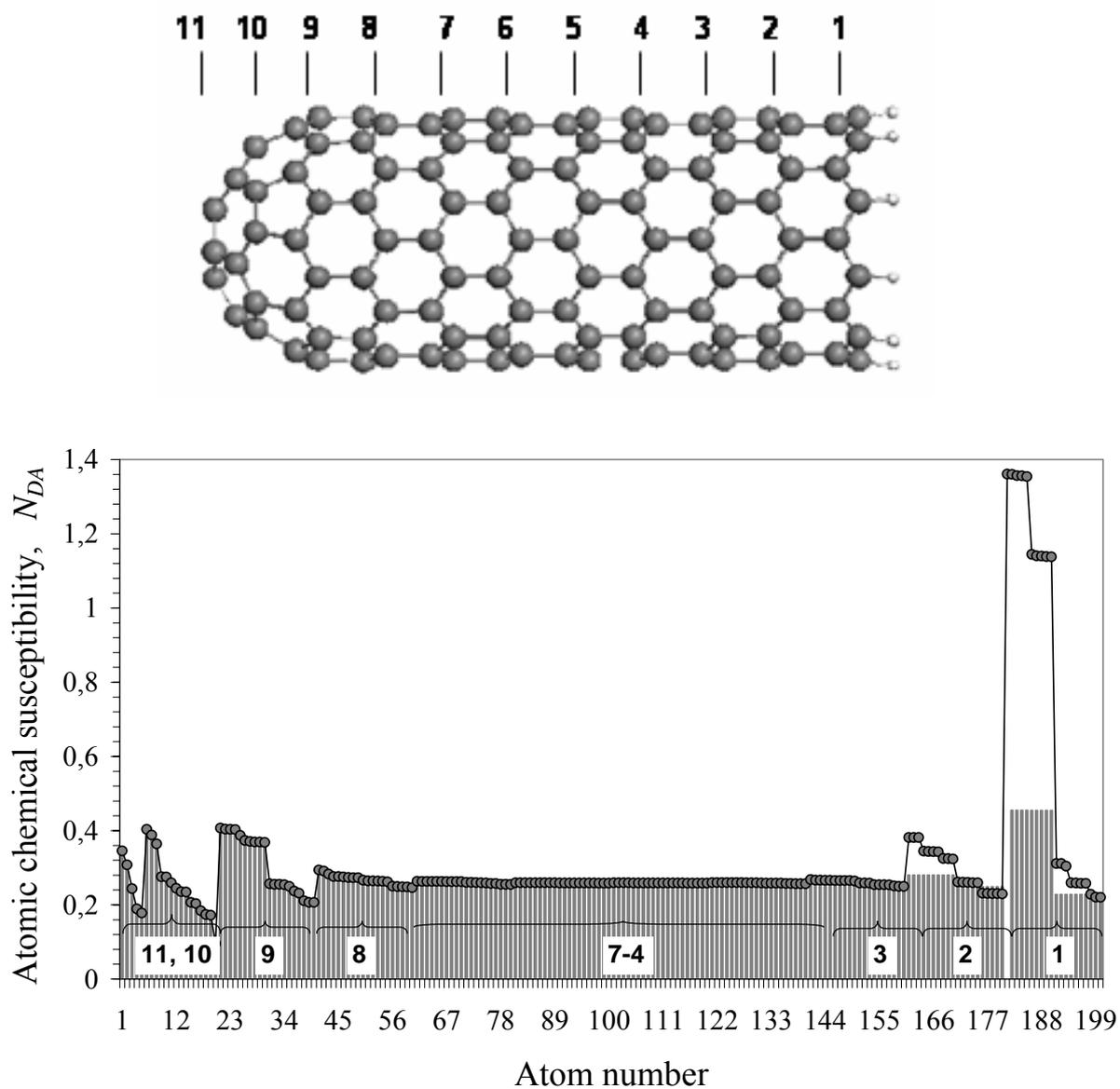

(b)

**Figure 2.** Atomic chemical susceptibility profiles of (10,10) SWCNT [13]. Hystogram corresponds to the tubes structure shown above. Curve plots the quantity when the tube right end is emptyfied. Atoms are numbered in succession from the cap towards the open end. Figures mark atom rows. UBS HF solution. Singlet state.

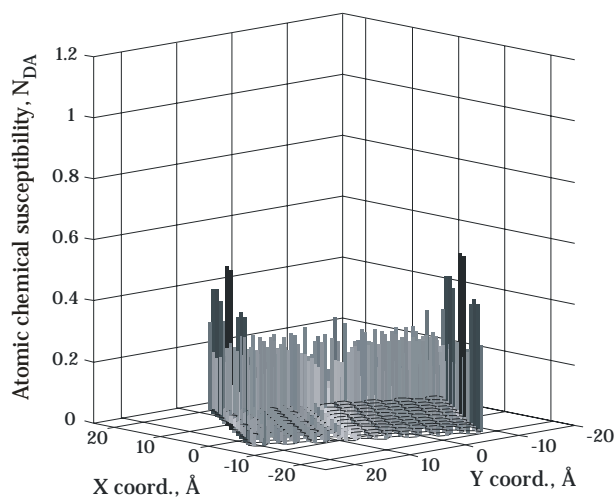 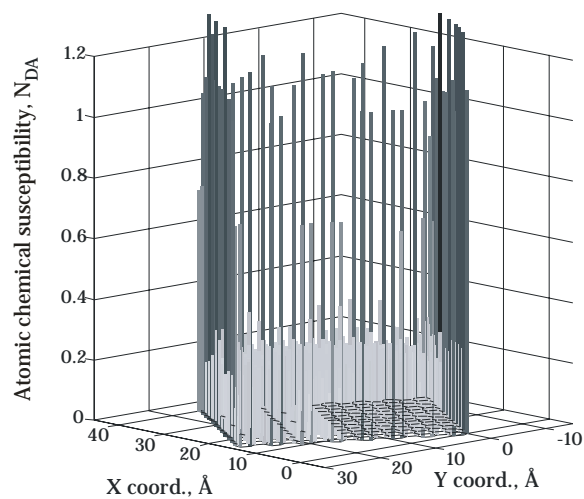

**Figure 3.** Distribution of atomic chemical susceptibility over atom of rectangular Ngr (15, 12) with hydrogen terminated (a) and empty (b) edges [14]. Figures in brackets correspond to the numbers of benzenoid units on the armchair and zigzag edges, respectively. UBS HF solution. Singlet state.

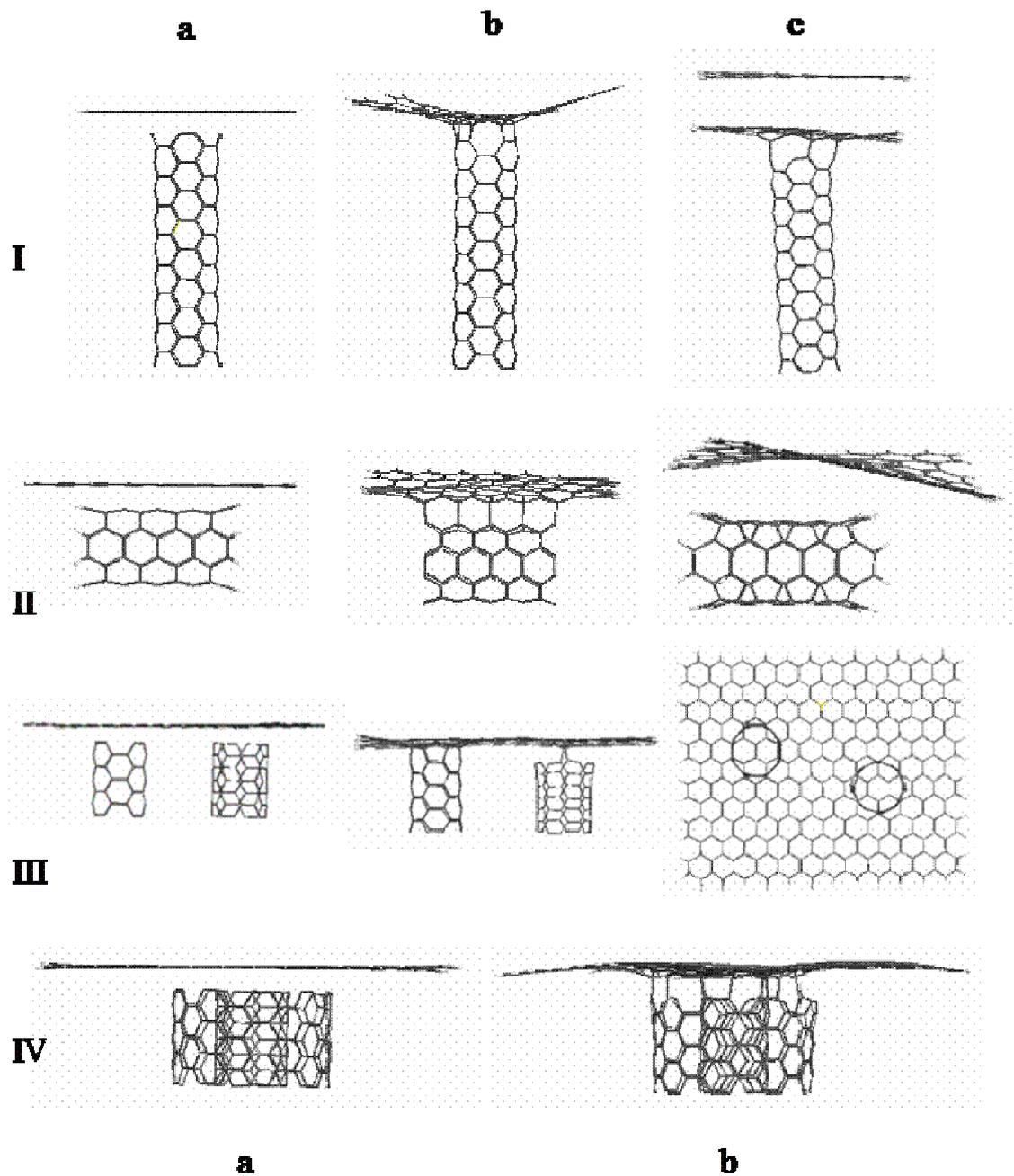

**Figure 4**. Hammer CNT- graphene composites. a) Starting and b), c) equilibrated compositions. **I**. (4,4) SWCNT with one (b) and two (c) NGrs (7,7). **II**. Ngr (7,7) and (4,4) SWCNT with both empty (b) and hydrogenated (c) ends. **III**. Two (4,4) SWCNTs and Ngr (11,12); side (a, b) and top (c) views. **IV**. Ngr (11,12) and (9,9+4,4) DWCNT.

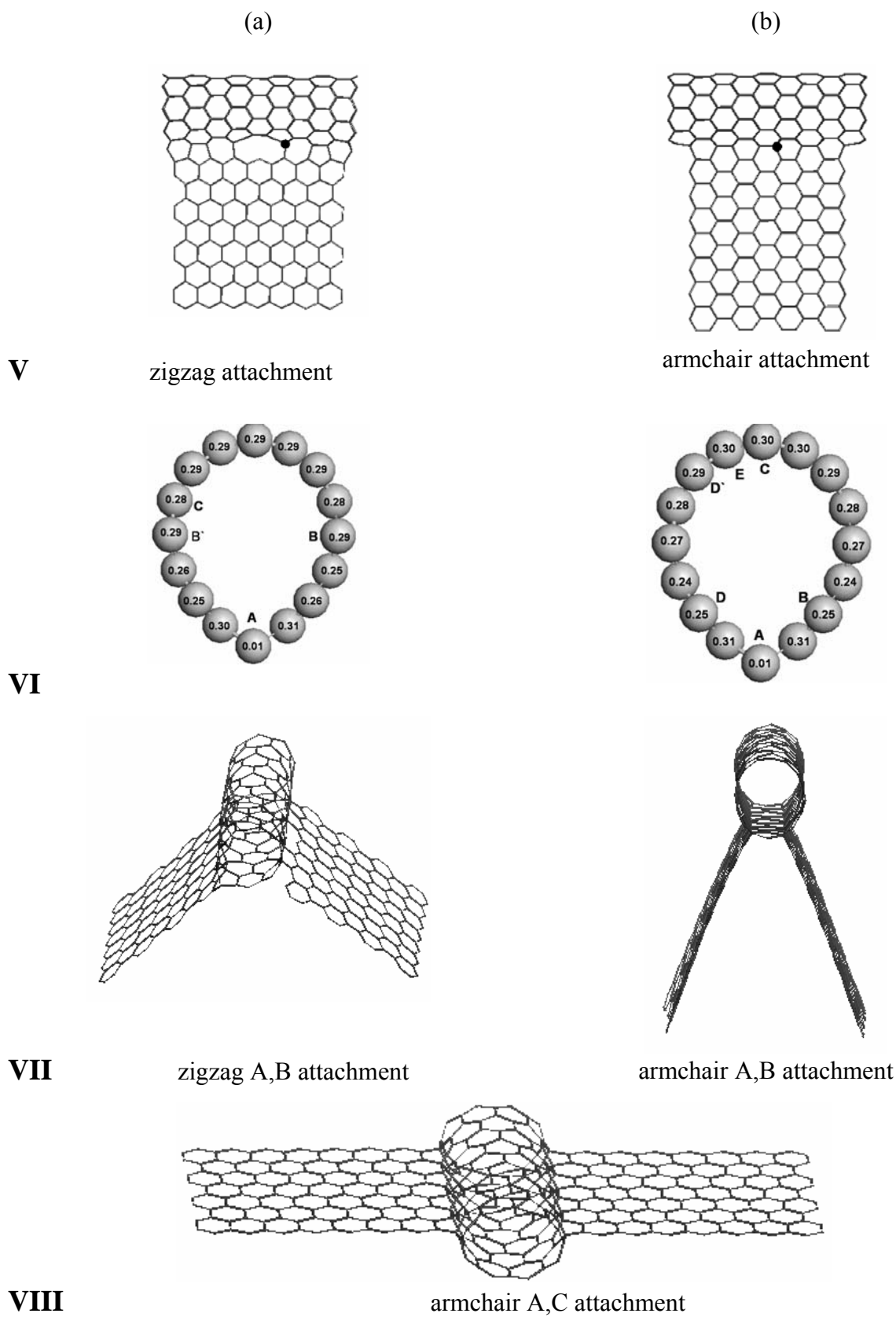

**Figure 5**. Cutting blade composites SWCNT (8,0) + 1 or 2 Ngr (7, 7). **V**. Equilibrium structures of monocomposites with interface along the line of atoms A at tube sidewall; zigzag (a) and armchair (b) attachments. **VI**. Distribution of $N_{DA}$ values over the tube cross section atoms in V*a* (a) and V*b* (b) composites, respectively. Ngr contacts the tube along the line involving atom A (A attachement). **VII**. Equilibrium structures of dicomposites related to A,B (VI*a* )zigzag (a) and A,B (VI*b*) armchair (b) attachments. **VIII**. The same as in **VII** but for A,C (VI*b*) armchair attachment (see text).

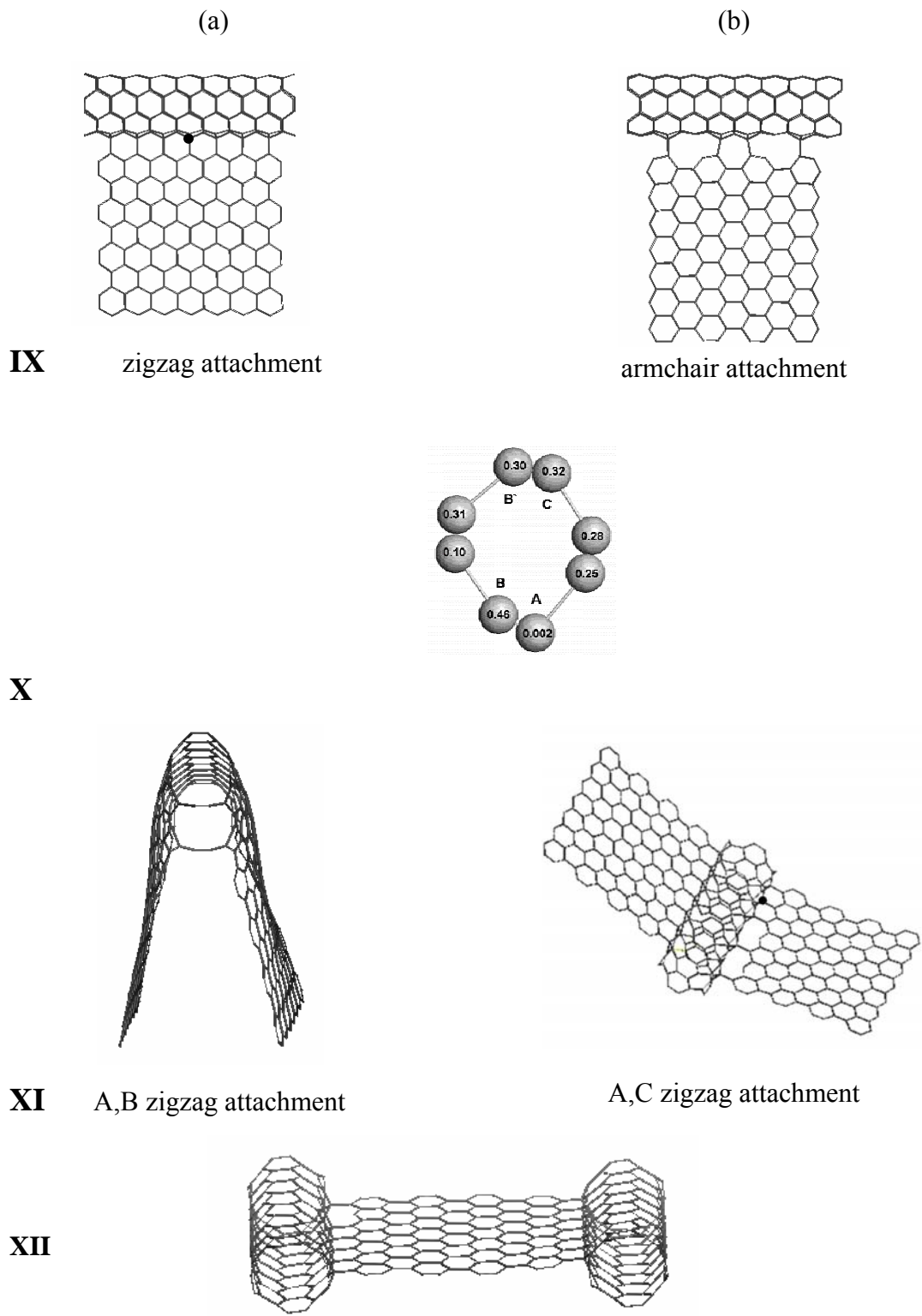

**Figure 6.** Cutting blade composites SWCNT (4,4) + 1 or 2 Ngr (7, 7). **IX**. Equilibrium structures of monocomposites with interface along the line of atoms **A** at tube sidewall; zigzag (a) and armchair (b) attachments. **X**. Distribution of $N_{DA}$ values over the tube cross section atoms in IX*a* composite. Ngr contacts the tube along the line involving atom A. **XI**. Equilibrium structures of dicomposites; A,B (a) and A,C (b) zigzag attachments. **XII**. Equilibrium structure of a "crandle" dicomposite; A-zigzag attachment of both tubes.